\documentclass[conference]{IEEEtran}
\IEEEoverridecommandlockouts
\usepackage[justification=centering]{caption}

\usepackage{cite}
\usepackage{amsmath,amssymb,amsfonts}
\usepackage{graphicx}
\usepackage{textcomp}
\usepackage{xcolor}

\usepackage{mathtools}
\usepackage{blkarray, bigstrut}
\usepackage{multirow, makecell}

\usepackage{algorithmic}
\usepackage{algorithm}
\usepackage{subcaption}

\usepackage{booktabs}

\usepackage{fontawesome}

\def\BibTeX{{\rm B\kern-.05em{\sc i\kern-.025em b}\kern-.08em
		T\kern-.1667em\lower.7ex\hbox{E}\kern-.125emX}}
\begin{document}
	
	\title{Towards Satellite Non-IID Imagery: A Spectral Clustering-Assisted Federated Learning Approach
	}
	
	\author{\IEEEauthorblockN{Luyao Zou$^\dag$, Yu Min Park$^\dag$, Chu Myaet Thwal$^\dag$, Yan Kyaw Tun$^\ddag$, Zhu Han$^*$, and Choong Seon Hong$^\dag$}
		\IEEEauthorblockA{$^\dag$ Department of Computer Science and Engineering, Kyung Hee University, Yongin-si, 17104, Republic of Korea \\
			$^\ddag$ Department of Electronic Systems, Aalborg University, A. C. Meyers Vænge 15, 2450 København \\
			* Department of Electrical and Computer Engineering, University of Houston, Houston, TX, 77004, USA}
 \thanks{This work has been submitted to the IEEE for possible publication. Copyright may be transferred without notice, after which this version may no longer be accessible.}
}


\maketitle
\begin{abstract}
	Low Earth orbit (LEO) satellites are capable of gathering abundant Earth observation data (EOD) to enable different Internet of Things (IoT) applications. However, to accomplish an effective EOD processing mechanism, it is imperative to investigate: 1) the challenge of processing the observed data without transmitting those large-size data to the ground because the connection between the satellites and the ground stations is intermittent, and 2) the challenge of processing the non-independent and identically distributed (non-IID) satellite data. In this paper, to cope with those challenges, we propose an orbit-based spectral clustering-assisted clustered federated self-knowledge distillation (OSC-FSKD) approach for each orbit of an LEO satellite constellation, which retains the advantage of FL that the observed data does not need to be sent to the ground. Specifically, we introduce normalized Laplacian-based spectral clustering (NLSC) into federated learning (FL) to create clustered FL in each round to address the challenge resulting from non-IID data. Particularly, NLSC is adopted to dynamically group clients into several clusters based on cosine similarities calculated by model updates. In addition, self-knowledge distillation is utilized to construct each local client, where the most recent updated local model is used to guide current local model training. Experiments demonstrate that the observation accuracy obtained by the proposed method is separately $1.01\times$, $2.15\times$, $1.10\times$, and $1.03\times$ higher than that of pFedSD, FedProx, FedAU, and FedALA approaches using the SAT4 dataset. The proposed method also shows superiority when using other datasets.
	
	

\end{abstract}


\vspace{-0.1cm}
\section{Introduction}
\par
Nowadays, with the fast growth of satellite technology, an increasing number of low-Earth orbit (LEO) satellites are being deployed in space \cite{Zhai_FedLEO_Offloading}. Those LEO satellites can be utilized to gather massive observed data to enable different applications in the Internet of Things (IoT) \cite{T_Xiong_Energy_Efficient_Federated_Learning_LEO}. \textit{However, owing to the intermittent nature of the connection between the satellites and the ground stations, sending massive observed data to the ground for processing still faces challenge \cite{T_Xiong_Energy_Efficient_Federated_Learning_LEO}.} To handle this issue, federated learning (FL), a distributed ML paradigm \cite{B_Matthiesen_FL_Satellite_Constellations}, can be introduced as the solution for satellite systems since it does not require data to leave the local side. 
\par
Regarding FL, it is constructed by a central server and multiple clients. The process of FL is as follows: Initially the global model generated by the central server will be distributed to each client for training. Then the updated local models will be sent to the central server for model aggregation such that to produce a new global model. FL has been adopted by several studies such as \cite{N_Razmi_Ground_FL_LEO, L_Wu_FedGSM_Gradient_Staleness, C_Yang_Communication_Progressive} to process data towards the satellite system. In \cite{N_Razmi_Ground_FL_LEO}, on the basis of FedAvg \cite{H_B_McMahan_Communication_Efficient_Learning_Deep_Networks}, a novel asynchronous FL procedure was proposed for LEO satellite constellations, where the satellite is treated as a client. In \cite{L_Wu_FedGSM_Gradient_Staleness}, FedGSM, an asynchronous FL approach, was proposed for LEO satellite constellations to process the local dataset of satellite, in which a compensation mechanism was introduced to reduce the gradient staleness. In \cite{C_Yang_Communication_Progressive}, an efficient satellite-ground FL approach (called SatelliteFL) was proposed, which guarantees that the satellite can finish training per round in each connection window.  Nevertheless, the satellite's route is relatively fixed, which can cause the individual satellites to capture fewer categories of Earth images, resulting in heterogeneous data (a.k.a. non-IID \cite{C_Li_FL_Soft_CLustering}) distribution among clients \cite{Zhai_FedLEO_Offloading}. \textit{Owing to the heterogeneous data, the performance of learning a shared global model in FL will be degraded \cite{C_Li_FL_Soft_CLustering}}.
\par
To deal with the aforementioned performance degradation issue, as per \cite{C_Li_FL_Soft_CLustering}, clustered FL (CFL) can be a possible way. Therefore, in this paper, an orbit-based spectral clustering-assisted clustered federated self-knowledge distillation (OSC-FSKD) approach is proposed to handle Non-IID satellite data. The main contributions of this paper are summarized as below:
\begin{itemize}
	\item We consider a multi-orbits-based LEO satellite constellation for performing an FL-based Earth observation application. The same as \cite{M_Elmahallawy_Stitching_Satellites_Edge_Pervasive_Efficient_LEO}, in each orbit, one satellite (called sink node) and multiple other satellites (called ordinary nodes) are considered.	
	\item We formulate a problem of maximizing the observation accuracy for the ordinary nodes in a single orbit under several constraints. To solve the formulated problem, the OSC-FSKD approach is proposed which is constructed based on normalized Laplacian-based spectral clustering (NLSC) algorithm, CFL, and self-knowledge distillation (SKD).
	\item
	Specifically, in each round, we perform NLSC in the sink node to make clients into several clusters according to the cosine similarities computed by local model updates. Then in the sink node, we adopt model aggregation (MA) in each cluster. From the second round, we employ SKD in each local client, where the knowledge from the previous round local model will be distilled to guide the training of current round local model. In the final round, we perform MA on the sink node across all participating clients to facilitate future applications. 
	\item Finally, evaluation by adopting convolutional neural network (CNN) and Swin Transformer \cite{Z_Liu_Swin_Transformer} as backbone network is provided to clarify the superiority of the proposed method by comparing it with baselines.
\end{itemize}

\par
The remaining of this paper is designed as: we provide the system model with the problem formulation in Section \ref{system_model_problem_formulation}. In Section \ref{Section_3_SC_FSKD_Approach}, the proposed solution is discussed. Section \ref{sec_4_Evaluation_Result_Analysis} and Section \ref{sec_5_conclusion} give the evaluation and conclusion, respectively.

\section{System Model and Problem Formulation}

\label{system_model_problem_formulation}
In this section, system model (e.g., network model) and the problem formulation are described for performing FL-based Earth observation task. 
\subsection{Network Model}
\label{section_2_Network_Model}
\begin{figure} [t!]
	\centering
	\includegraphics[scale = .43]{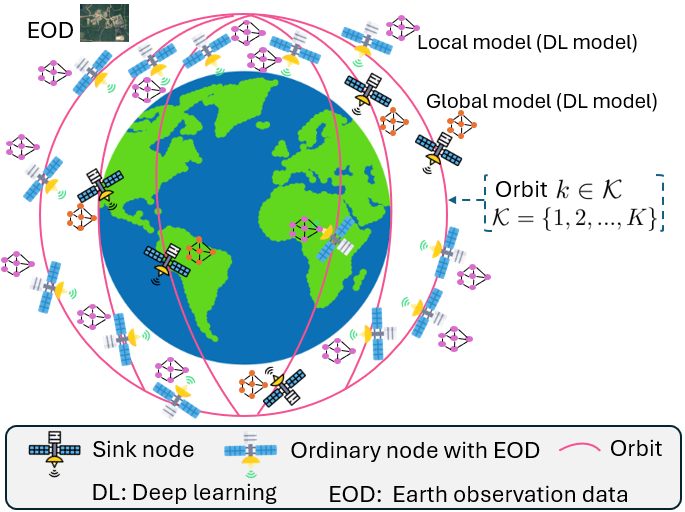}
	\caption{System model for LEO satellite constellation.}
	\label{fig_SystemModel} 
	\vspace{-0.55cm}
\end{figure}

Consider a LEO satellite constellation (SateCon) for FL-based earth observation, shown in Fig. \ref{fig_SystemModel}. This LEO SateCon has a set $\mathcal{K}= \{1, 2, ..., K\}$ of orbits.
Inspired by \cite{M_Elmahallawy_Stitching_Satellites_Edge_Pervasive_Efficient_LEO} which considers using one satellite (called sink node) for model aggregation in the FL process to generate the global model, and considers using other satellites to perform the local model update process, in this work, for each orbit $k\in \mathcal{K}$, we consider it contains a sink node for model aggregation (described in Section \ref{Section_3_SC_FSKD_Approach}), and a set $\mathcal{S}_{k} = \{1, 2, ..., S_{k}\}$ of $S_{k}$ other satellites (called ordinary nodes for simplicity) for local model update. It is noteworthy that the choice of sink node is out of the scope of this paper. The ordinary nodes are utilized to capture the Earth's observation images and perform the local Earth observation task via deep learning (DL). Same as \cite{M_Elmahallawy_Stitching_Satellites_Edge_Pervasive_Efficient_LEO}, we consider the images will be captured while satellites orbiting the Earth. Identical to \cite{N_Razmi_On_Board_FL_Satellite_Clusters_ISL}, we consider the satellites in the same orbit are allocated unique identifiers (IDs). The set of unique IDs for the satellites in orbit $k$ is denoted as $\mathcal{I}_k = \{0, 1, 2, ..., I_k\}$, where the ID of sink node is $0$ and the IDs of ordinary nodes are from $1$ to $I_k$. Further, communications between satellites in the considered LEO SateCon are carried out via inter-satellite links (ISLs) \cite{Leyva_Mayorga_Inter_Plane_IS_Connectivity_Dense}. Same as \cite{X_Zhang_Energy_Efficient_Computation_Satellite}, we assume laser ISL is adopted.   	




\subsection{Satellites Communication Model}
As per \cite{F_Zhou_Decentralized_Satellite_FL_Intra_Inter}, the communication link between satellites is able to be established only under the situation that the Earth does not block the Line-of-Sight (LoS). We can give the condition for constructing the feasible inter-satellite communication as $\psi(s_1, s_2) < \psi_{th}(s_1, s_2)$ \cite{Leyva_Mayorga_Inter_Plane_IS_Connectivity_Dense}, where $s_1$ and $s_2$ denote two satellites belongs to $\mathcal{S}_{k}$. $\psi(s_1, s_2)$ represents the distance between two satellites, while $\psi_{th}(s_1, s_2)$ indicates the distance threshold that can be given as follows \cite{A_Zhou_Two_Layer_FL_LEO_IoT}:
\begin{subequations}\label{Opt_1}
	\vspace{-0.1cm}
	\setlength{\abovedisplayskip}{3pt}
	\setlength{\belowdisplayskip}{3pt}
	\begin{align}
		&\label{Opt_1:const1}
		\psi_{th}(s_1, s_2) = \notag \\ & \sqrt{(h_{s_1} + R_E)^2 - {R_E}^2} + \sqrt{(h_{s_2} + R_E)^2 - {R_E}^2}, \tag{1}
	\end{align}
\end{subequations} 
where $h_{s_1}$ and $h_{s_2}$ are the altitudes of satellite $s_1$ and $s_2$, respectively \cite{A_Zhou_Two_Layer_FL_LEO_IoT}. $R_E = 6, 371$ km \cite{M_Alsenwi_Flexible_Synchronous_LEO} is the Earth radius.

\subsection{Deep Learning Model for Satellite Imagery Observation}
\label{section_2_C_DL_satellite_imagery}
To observe the Earth, observing the image data collected by the satellites, through the DL model, is considered in this work. Particularly, we consider deploying each DL model in each satellite, where the DL model in each ordinary node is called the local model, and that on the sink node is dubbed the global model, as shown in Fig. \ref{fig_SystemModel}. 
\par
In general, the DL model is formed by 1) the representation layers, and 2) the decision layers \cite{Y_Tan_FedProto_Heterogeneous} (a.k.a. classifier). For an ordinary node of an orbit, $s \in \mathcal{S}_{k}$, the set of images owned by $s$ can be represented as $\mathcal{X}_s = \{x_s^1, x_s^2, ..., x_s^M\}$, and the set of labels corresponding to the image set $\mathcal{X}_s$ is denoted as $\mathcal{Y}_s = \{y_s^1, y_s^2, ..., y_s^M\}$. Here, $M$ is the overall number of images owned by each ordinary node. In addition, we denote the representation layers as $\xi_s(\cdot)$ and the classifier as $\Psi_s(\cdot)$. Hence, for an image, $x_s^m \in \mathcal{X}_s$, the extracted features from the representation layers can be given as $\delta_s^m = \xi_s(x_s^m)$, while the final observed result, $\widehat{y_s^m}$, can be defined by:
\begin{subequations}\label{Opt_2}
	\vspace{-0.1cm}
	\setlength{\abovedisplayskip}{3pt}
	\setlength{\belowdisplayskip}{3pt}
	\begin{align}
		&\label{Opt_2:const1}
		\hspace{-0.6cm} \widehat{y_s^m} = (\Psi_s \circ \xi_s)(x_s^m) \notag = \Psi_s\Big(\xi_s(x_s^m)\Big) = \Psi_s(\delta_s^m), \tag{2}
	\end{align}
\end{subequations}
where in mathematics, $\circ$ is the composition of functions. 

\vspace{-0.15cm}
\subsection{Problem Formulation}
\vspace{-0.15cm}
\label{Section_2_Problem_Formulation}
In this work, based on the observed Earth image data, we consider the observation task via each local DL model, aiming to maximize the observation accuracy (OA) of all ordinary nodes for a orbit $k$, as only the ordinary nodes are used to capture the image data. To be specific, OA is defined as the quotient of the number of correctly observed Earth data and the total number of Earth observation data. Next, we will illustrate the calculation process regarding OA in detail.

\par
Consider the image data of each ordinary node $s \in \mathcal{S}_{k}$ belongs to total $C_s$ categories. For convenience, let $\mathcal{C}_s = \{1, 2, ..., c\}$ denote a set of categories owned by each ordinary node $s \in \mathcal{S}_{k}$. For a given image $x_s^m \in \mathcal{X}_s$ of satellite $s$ in orbit $k$, to indicate whether the observed result for $x_s^m$ is class $c \in \mathcal{C}_s$ or not, for simplicity, a binary decision variable for each class $c$ is introduced, which is defined as:
\begin{subequations}\label{Opt_3}
	\setlength{\abovedisplayskip}{3pt}
	\setlength{\belowdisplayskip}{3pt}
	\begin{align}
		\hspace{-0.3cm}	\alpha_s^{c}(x_s^m) = \left\{  
		\begin{array}{rcl}
			1, & \textrm{if $\widehat{y_s^m} = c$ \& $y_s^m=c$}, \\
			0,&  \hspace{-1.65cm}  \textrm{otherwise.}  \\
		\end{array} \right. \hspace{-0.1cm} \tag{3}
	\end{align}
\end{subequations} 
In \eqref{Opt_3}, with given input image $x_s^m$, if the true label is $c$, and the observed result is also $c$, then $\alpha_s^{c}(x_s^m) = 1$, otherwise, $\alpha_s^{c}(x_s^m) = 0$. Namely, if $\alpha_s^{c}(x_s^m) = 1$, the observed result is correct. Thus, the correctly observed data of all participating ordinary nodes in orbit $k$, $\Lambda_k$, can be defined as:
\begin{subequations}\label{Opt_4}
	\begin{align}
		&\label{Opt_4:const1}
		\Lambda_k = \sum_{s=1}^{|\mathcal{S}_{k}|}\sum_{m=1}^{M}\alpha_s^{c}(x_s^m) , \tag{4}
	\end{align}
\end{subequations}
where $|\mathcal{S}_{k}|$ is the overall number of ordinary nodes in orbit $k$.
\par	
In the real world, satellites can become failures in space because of different factors (e.g., natural shocks and unexpected situations) \cite{Y_Cao_Detection_Method_Image_Enhancement_R_CNN}. Therefore, it is hard to assume all satellites can always provide service (e.g., training). Accordingly, it is reasonable to consider the availability of satellites. Specifically, as aforementioned, we adopt sink node to perform model aggregation process for FL-based Earth observation task. Thus, if the sink node is failure, the FL-based task can not be executed. For simplicity, we ignore the consideration of sink node failure in this work. For each ordinary node $s \in \mathcal{S}_k$, we adopt a binary decision variable to indicate its availability (denoted as $\beta_{s}$) for convenience, which is given as below:
\begin{subequations}\label{Opt_5}
	\setlength{\abovedisplayskip}{3pt}
	\setlength{\belowdisplayskip}{3pt}
	\begin{align}
		\beta_{s} = \left\{  
		\begin{array}{rcl}
			1, & \textrm{if ordinary node $s$ is not failure}, \\
			0,&  \hspace{-3.2cm }\textrm{otherwise.}  \\
		\end{array} \right.  \tag{5}
	\end{align}
\end{subequations} 
In \eqref{Opt_5}, if $\beta_s=1$, the ordinary node $s$ is available to provide service. Otherwise, $\beta_s=0$. Accordingly, the overall number of Earth observation data of all participating ordinary nodes $\Omega_k$ can be given as $\Omega_k = \sum_{s=1}^{|\mathcal{S}_{k}|} \beta_s M_s$. Here, $M_s$ is the total number of image observed by the  ordinary node $s$.
\par
As a result, given the availability $\beta_{s}$, i.e., OA for the ordinary nodes in an orbit, can be given as:
\begin{subequations}\label{Opt_6}
	\setlength{\abovedisplayskip}{2pt}
	\setlength{\belowdisplayskip}{2pt}
	\begin{align}
		\underset{\boldsymbol{\alpha}, \boldsymbol{\beta}} \max 
		&\; \quad
		\frac{\sum_{s=1}^{|\mathcal{S}_{k}|}\sum_{m=1}^{M} \beta_s \alpha_s^{c}(x_s^m)}{\Omega_k},  \tag{6} \\
		\text{s.t.} \qquad \qquad \qquad
		&\label{Opt_6:const1} 
		\hspace{-1.1 cm} \psi(s_1, s_2) < \psi_{th}(s_1, s_2), \\
		&\label{Opt_6:const2} 
		\hspace{-1.1 cm} \sum_{s=1}^{|\mathcal{S}_{k}|} \beta_s \leqslant |\mathcal{S}_{k}|, \beta_s \in \{0, 1\}\\
		&\label{Opt_6:const3} 
		\hspace{-1.1 cm} \alpha_s^{c}(x_s^m) \in \{0, 1\}, x_s^m \in \mathcal{X}_s, \\
		&\label{Opt_6:const4}
		\hspace{-1.1 cm} \boldsymbol{\alpha} = \{\alpha_s^{c}(x_s^m), s \in \mathcal{S}_k, c \in \mathcal{C}_s, x_s^m \in \mathcal{X}_s\}, \\
		&\label{Opt_6:const5}
		\hspace{-1.1 cm} \boldsymbol{\beta} = \{\beta_s, s \in \mathcal{S}_k\},
	\end{align}	
\end{subequations} 
where \eqref{Opt_6:const1} ensures the communication link can be constructed between satellites. \eqref{Opt_6:const2} assures that the total available ordinary nodes in orbit $k$ do not exceed the overall number of ordinary nodes in orbit $k$, where $\beta_s$ is a binary decision variable. \eqref{Opt_6:const3} guarantees $\alpha_s^{c}(x_s^m)$ is a binary decision variable. \eqref{Opt_6:const4} and \eqref{Opt_6:const5} are the decision $\boldsymbol{\alpha}$ and $\boldsymbol{\beta}$, respectively. They are separately related to the binary decision $\alpha_s^{c}(x_s^m)$ and $\beta_s$.

\vspace{-0.1cm}
\section{Proposed SC-FSKD Approach for Non-IID Satellite Imagery}
\label{Section_3_SC_FSKD_Approach}
\begin{figure*} [htpb]
	\centering
	\includegraphics[scale = .60]{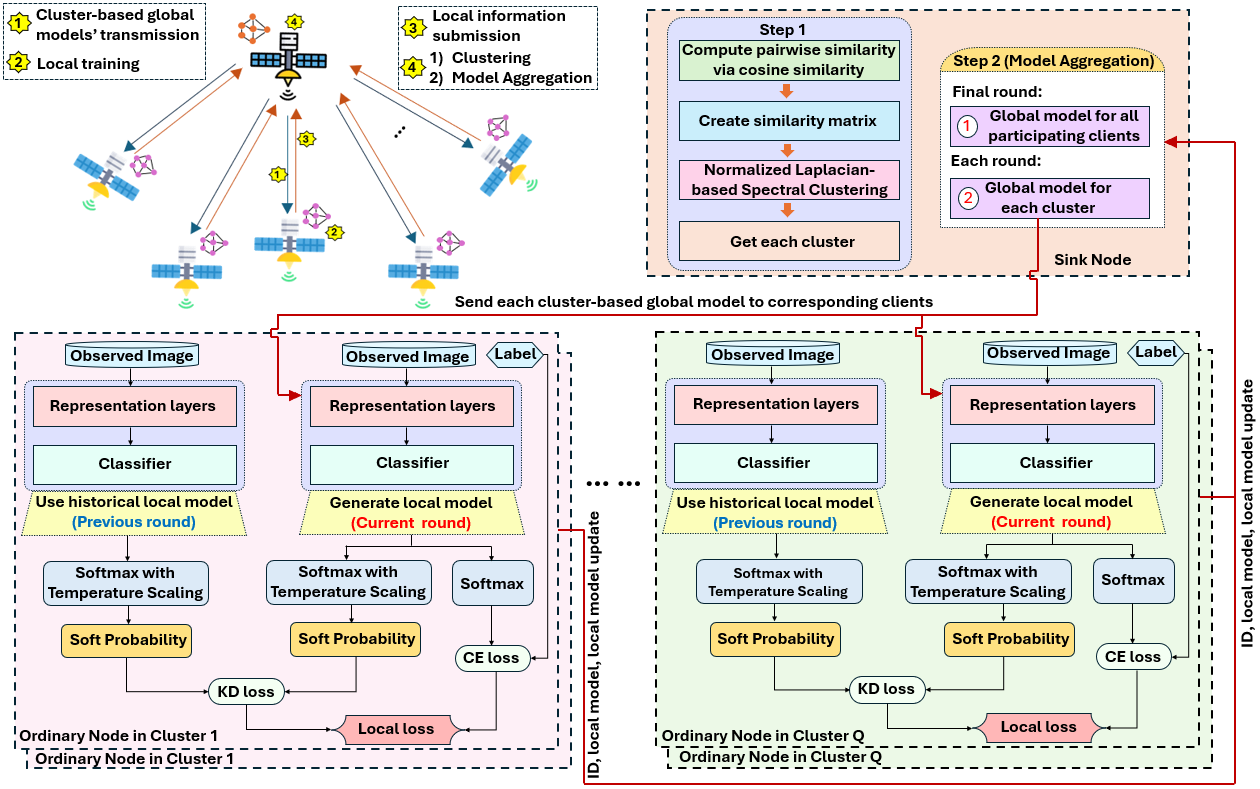}
	\caption{Architecture of the proposed OSC-FSKD approach for satellite imagery observation in each orbit.}
	\label{fig_solution} 
	\vspace{-0.45cm}
\end{figure*}
\par  	
To devise the solution to solve \eqref{Opt_6}, we propose an orbit-based spectral clustering-assisted clustered federated self-knowledge distillation (OSC-FSKD) approach, which is designed based on normalized Laplacian-based spectral clustering (NLSC), FL and SKD. Concretely, for FL,  \texttt{FedAvg} \cite{H_B_McMahan_Communication_Efficient_Learning_Deep_Networks} is adopted by the proposed OSC-FSKD approach, and we consider performing the proposed OSC-FSKD approach in each orbit. Fig. \ref{fig_solution} shows the architecture of the proposed OSC-FSKD approach for satellite imagery observation in each orbit. Specifically, each ordinary node in an orbit $k$ is regarded as each local client, while the sink node in orbit $k$ is regarded as the server for the proposed OSC-FSKD approach. In addition, the communication between the sink node and ordinary node may be not available (i.e., constraint \eqref{Opt_6:const1} violation). To solve this issue, inspired by \cite{M_Elmahallawy_Stitching_Satellites_Edge_Pervasive_Efficient_LEO} which adopts the model relaying process (MRP) in terms of forwarding the model to next-hop neighbors and the neighbors will continue doing the same process, we assume the MRP is available in this work to make sure the model can be sent to the target node. Next, we will give the details of the proposed solution.
\vspace{-0.2cm}
\subsection{Local Client Design via SKD (Ordinary Node)}
\vspace{-0.1cm}
\label{Section_3_Local_Client_Design_Satellite_Imagery}
To construct each local client (i.e., ordinary node), we adopt SKD \cite{H_Jin_Personalized_Edge_FSKD} in each local client. Specifically, each client is formed by a DL network that contains representation layers and classifier, as mentioned in Section \ref{section_2_C_DL_satellite_imagery}. The DL network is utilized to generate the local model. Based on \cite{H_Jin_Personalized_Edge_FSKD}, we consider transferring the knowledge from the most recently updated past model to the current local model, as shown in Fig. \ref{fig_solution}. To be more specific, the local model obtained in the previous round will be treated as the teacher model, and its knowledge will be distilled to guide the local model training in current round (regarded as student model). It is worth mentioning that the SKD will be performed from the second round because of the necessity of using the recently updated past model.

\par
Since the soft probabilities of the teacher network can be leveraged to supervise the student network \cite{J_H_Cho_Efficacy_KD}, we introduce soft probabilities that can be computed by using \textit{Softmax with Temperature scaling} \cite{G_Hinton_Distilling_Knowledge_NN} (softmaxT). Hence, in this article, softmaxT is employed in the considered SKD, where the output of the classifier will be converted to the soft probability. To be specific, given an image $x_s^m \in \mathcal{X}_s$ of an ordinary node $s \in \mathcal{S}_k$ in orbit $k$, using the latest updated past model $\omega_s^{t-1}$ (i.e., teacher model), the output of classifier generated by the teacher model per satellite $s$ can be represented as $z_s(x_s^m) = \{z_s^1(x_s^m), z_s^2(x_s^m), ..., z_s^c(x_s^m)\}$. Here, $t-1$ and $t$ denote the $(t-1)^\textrm{th}$ and the $t^\textrm{th}$ round for the proposed solution, while $c \in \mathcal{C}_s$ indicates the class $c$. Therefore, the soft probability obtained by teacher model, denoted as $\zeta_s^\textrm{c}(x_s^m)$, can be given as $\zeta_s^\textrm{c}(x_s^m) = \frac{e^{z_s^c(x_s^m)/\tau}}{\sum_{o} e^{z_s^o(x_s^m)/\tau}}$ \cite{G_Hinton_Distilling_Knowledge_NN},
where $z_s^o(x_s^m)$ is the $o^\textrm{th}$ elements of $z_s(x_s^m)$, and $\tau$ indicates the temperature. The soft probability of the current trained local model (i.e., student model), $\widehat{\zeta_s^c}(x_s^m)$, can be gained in a similar manner. Consequently, the KD loss, a part of the local client's loss function, can be computed by \cite{B_Peng_Correlation_Congruence_ICCV}:
\begin{subequations}\label{Opt_7}
	\setlength{\abovedisplayskip}{3pt}
	\setlength{\belowdisplayskip}{3pt}
	\begin{align}
		&\label{Opt_7:const1}
		\mathcal{L}_s^\textrm{KD} =  \frac{1}{|\mathcal{X}_s|}\sum_{m=1}^{|\mathcal{X}_s|}\tau^2KL\Big(\zeta_s^c(x_s^m), \widehat{\zeta_s^c}(x_s^m)\Big), \tag {7}
	\end{align}	
\end{subequations}
where $KL(\cdot, \cdot)$ is the Kullback-Leibler (KL) divergence.
\par
\textbf{Local Objective } To maximize observation correctness of an orbit, i.e., \eqref{Opt_6}, we consider minimizing the local loss function of each participating ordinary node. In particular, the local loss function is defined as the cross entropy (CE) loss (denoted as $\mathcal{L}_s^\textrm{CE}$) in the first round, as SKD is not performed yet, and is defined as the linear combination of the KD loss \eqref{Opt_7} and the CE loss in other rounds, which is given as below:
\begin{subequations}\label{Opt_8}
	\vspace{-0.1cm}
	\setlength{\abovedisplayskip}{3pt}
	\setlength{\belowdisplayskip}{3pt}
	\begin{align}
		\mathcal{L}_s^\textrm{loss} = \left\{  
		\begin{array}{rcl}
			\mathcal{L}_s^\textrm{CE}, & \textrm{if $t=1$}, \\
			\varphi \mathcal{L}_s^\textrm{CE} + (1-\varphi) \mathcal{L}_s^\textrm{KD},& \hspace{0.21cm} \textrm{otherwise.}  \\
		\end{array} \right.  \tag{8}
	\end{align}
\end{subequations} 
where $\varphi$ is a hyperparameter. $t=1$ means the first round. Hence, the local objective can be defined as:	
\begin{subequations}\label{Opt_9}
	\setlength{\abovedisplayskip}{3pt}
	\setlength{\belowdisplayskip}{3pt}
	\begin{align}
		\min 
		&\; \mathcal{L}_s^\textrm{loss}. \tag {9}
	\end{align}	
\end{subequations}

\subsection{Information Submission from Ordinary Nodes to Sink Node}
\par
For the standard FL, after training, the client will submit the current local model to the server for model aggregation \cite{S_I_Nanayakkara_Global_Aggregation_FL}. Different from this point, in this work, for each participating ordinary node (i.e., $\beta_s = 1$), we consider sending the information including ID ($i_{n} \in \mathcal{I}_k\backslash\{0\}$), local model and local model update of each participating client to the sink node for clustering and model aggregation (illustrated in Section \ref{Section_3_NLSC_CMA}). Here, $\mathcal{I}_k\backslash\{0\}$ means the set $\mathcal{I}_k$ but the element $0$ is removed, as the element $0$ is the ID of sink node. Namely, $\mathcal{I}_k\backslash\{0\}$ is the set of IDs for the ordinary nodes of orbit $k$. 
\par
To be specific, for a participating ordinary node $s \in \mathcal{S}_k$ in orbit $k$, let its local model before update and after update at round $t$ are denoted as $\omega_s^\textrm{before}$ and $\omega_s^t$, respectively. It is noteworthy that $\omega_s^\textrm{before}$ is the obtained global model for the cluster of the previous round that the ordinary node $s$ in, which will be illustrated in Section \ref{Subsection_3_Cluster_Model_Aggregation}. Thus, we can get the local model update of the ordinary node $s$ in a round as $\Delta\omega_s^t = \omega_s^t - \omega_s^\textrm{before}$. In summary, in each round $t$, we will send $\{i_{n}, \omega_s^t, \Delta\omega_s^t\}$ of each ordinary node $s$ to the sink node. Notably, the ID will be used by the sink node to distinguish whether a client belongs to a generated cluster. Thereby, the sink node can know whether should send a generated cluster-based global model to that client for next-round training. The cluster generation process will be illustrated in the next.

\subsection{Normalized Laplacian-based Spectral Clustering for Grouping Clients and Model Aggregation on the Sink Node}
\label{Section_3_NLSC_CMA}
\par
As the clustered federated learning (CFL) is possible to handle the performance degradation issue caused by using the heterogeneous data to learn a shared global model in FL \cite{C_Li_FL_Soft_CLustering}, CFL is considered in this work. Specifically, CFL is the method to partition clients into various clusters such that to train the cluster-level models \cite{Y_L_Tun_Contrastive_CFL_Heterogeneous_Data}. In this article, in each round, to make the participating ordinary nodes into several clusters for each orbit $k$, a \textbf{normalized Laplacian-based spectral clustering (NLSC)} technique is employed. Then, in each cluster, model aggregation (MA) will be performed in each round. Lastly, to facilitate future application, we perform MA on all participating clients in the final round to generate a global model. Next, we will explain each step in detail.
\subsubsection{\textbf{NLSC for clustering clients}}
NLSC will be performed on the sink node. Concretely, considering there are total $f$ participating clients (i.e., ordinary node with $\beta_s=1$) in orbit $k$ in each round. To group the clients, the cosine similarity calculated by model updates \cite{M_Duan_FedGroup} is adopted in this work. Specifically, we consider clustering those participating clients by exploring a set $\mathcal{G}=\{G_{1}, G_{2}, ..., G_{f}\}$ of cosine similarities computed by using the model updates. Each element $G_{f}$ is the cosine similarity between the model update of the $f^\textrm{th}$ participating client and each participating client. Notably, the cosine similarity between a model update and itself is equal to $1$. Thus, the element $G_{1}$ can be denoted as $G_{1} = \{1, g_{1}^2, ..., g_{1}^{f}\}$, similarly, $G_{2}$ can be denoted as $G_{2} = \{g_{2}^1, 1, ..., g_{2}^{f}\}$, and $G_{f}$ can be denoted as $G_{f} = \{g_{f}^1, g_{f}^2, ..., 1\}$, by analogy. Particularly, $g_{f_1}^{f_2}$ is the cosine similarity between the model updates of two participating clients $f_1$ and $f_2$, which can be defined as \cite{M_Duan_FedGroup}:
\begin{subequations}\label{Opt_10}
	\vspace{-0.25cm}
	\setlength{\abovedisplayskip}{3pt}
	\setlength{\belowdisplayskip}{3pt}
	\begin{align}
		&\label{Opt_10:const1}
		g_{f_1}^{f_2} \triangleq \frac{<\Delta\omega_{f_1}^t, \Delta\omega_{f_2}^t>}{\parallel \Delta\omega_{f_1}^t \parallel \parallel \Delta\omega_{f_2}^t \parallel}, \tag {10}
	\end{align}	
\end{subequations}
where $<\cdot, \cdot>$ denotes the inner product \cite{S_Zhu_Scaling_Up_TopK}, while $\parallel\cdot\parallel$ represents the $L_2$ norm \cite{S_Zhu_Scaling_Up_TopK}.	
	\par
	For NLSC approach, according to \cite{L_Zou_Clustering_Serverless_Edge_Computing_FL}, it is necessary to construct an undirected graph $\mathcal{B}$ according to $\mathcal{G}$, where each $G_{f}$ is the vertex of $\mathcal{B}$. As for the edges of $\mathcal{B}$, they are weighted based on the similarity measure between each pair of vertexes $(G_{f_1}, G_{f_2})$ \cite{G_Sun_What_and_How}. Accordingly, we can create a weighted affinity matrix $\widetilde{W} \in \mathbb{R}^{f \times f}$ with each element $\widetilde{w}_{f_1, f_2}$. Particularly, $\widetilde{w}_{f_1, f_2}$ is the similarity measure between $G_{f_1}$ and $G_{f_2}$, which can be computed by \cite{Y_Wang_Spectral_Clustering_Multiple_Manifolds}:
	\begin{subequations}\label{Opt_11}
		\setlength{\abovedisplayskip}{3pt}
		\setlength{\belowdisplayskip}{3pt}
		\begin{align}
			&\label{Opt_11:const1}
			\widetilde{w}_{f_1, f_2} = \left\{  
			\begin{array}{rcl}
				\textrm{exp}(\frac{-\left \|\mathcal{G}_{f_1} -\mathcal{G}_{f_2} \right\|^2}{2\hat{\sigma}^2}), &\textrm{if } f_1 \neq f_2, \\
				0,& \hspace{-0.1cm} \textrm{otherwise,}  \\
			\end{array} \right. \tag{11}
		\end{align}  	
	\end{subequations}
	where $\hat{\sigma}$ indicates a free parameter, while $\parallel\cdot\parallel$ denotes the Euclidean norm \cite{Y_Wang_Spectral_Clustering_Multiple_Manifolds} (a.k.a. $L_2$ norm \cite{D_Dadush_Enumerative_Lattice_Algorithms_M_ellipsoid}), same as \eqref{Opt_10}. Then the created weighted affinity matrix $\widetilde{W}$ will be used to obtain the normalized Laplacian $A$. As per \cite{G_Sun_What_and_How}, $A = P^{-\frac{1}{2}}\widetilde{W}P^{-\frac{1}{2}}$, where $P$ denotes the diagonal matrix with each diagonal element $P_{f_1f_1}$ given as $P_{f_1f_1}=\sum_{f_2=1}^f \widetilde{w}_{f_1, f_2}, \forall f_1$ \cite{G_Sun_What_and_How}.
	\begin{figure}[t!]
		\vspace{-0.25cm}
		\begin{algorithm}[H]	
			\renewcommand{\algorithmicrequire}{\textbf{Input:}}
			\renewcommand{\algorithmicensure}{\textbf{Output:}}
			\caption{Normalized Laplacian-based Spectral Clustering Algorithm in Each Round for Clusters Design}
			\label{alg_1}
				\begin{algorithmic}[1]
					\REQUIRE $\mathcal{G}=\{G_{1}, G_{2}, ..., G_{f}\}$, cluster number: $Q$
					\ENSURE Clients of $Q$ clusters
					\STATE Construct a weighted affinity matrix $\widetilde{W} \in \mathbb{R}^{f \times f}$ by calculating $\widetilde{w}_{f_1, f_2}$ based on \eqref{Opt_11}
					\STATE Compute $P_{f_1f_1}$ by $P_{f_1f_1}=\sum_{f_2=1}^f \widetilde{w}_{f_1, f_2}, \forall f_1$ \cite{G_Sun_What_and_How} to create diagonal matrix $P$
					\STATE Obtain normalized Laplacian $A$ by $A = P^{-\frac{1}{2}}\widetilde{W}P^{-\frac{1}{2}}$
					\STATE Find $Q$ largest eigenvectors of $A$: $\widehat{O}_1, \widehat{O}_2, \dots, \widehat{O}_Q$
					\STATE Build a matrix $\Upsilon$ including the eigenvectors as columns: $\Upsilon = [\widehat{O}_1 \widehat{O}_2 \dots \widehat{O}_Q] \in \mathbb{R}^{f\times Q}$
					\STATE Build a matrix $D$ by normalizing the rows of $\Upsilon$ (i.e., $D_{ij}=\frac{\Upsilon_{ij}}{(\sum_{j}\Upsilon_{ij}^2)^{1/2}}$ \cite{A_Y_Ng_SC_Analysis})
					\STATE Regard each row of $D$ as a data point, and then make those data points into clusters via KMeans algorithm
					\STATE Assign $G_{f} \in \mathcal{G}$ to the cluster to which the $f^{th}$ row of $D$ belongs
					\STATE Get the corresponding clients for each cluster.  
					\STATE \textbf{return} Clients of each cluster  
				\end{algorithmic}  
			\end{algorithm} 
			\vspace{-0.9cm}
		\end{figure}
		\par
		The elaborate process of NLSC is listed in Algorithm \ref{alg_1}. Concretely, in line $1$, a weighted affinity matrix $\widetilde{W}$ is created via computing $\widetilde{w}_{f_1, f_2}$ as per \eqref{Opt_11}. Then we create diagonal matrix $P$ in line $2$, compute normalized Laplacian $A$ in line $3$ and obtain $Q$ largest eigenvectors of $A$ in line $4$. Afterward, the obtained $Q$ largest eigenvectors will be used to build a matrix $\Upsilon$ in line $5$. In line $6$, $\Upsilon$ will be used to construct another matrix $D$ through normalizing the rows of $\Upsilon$. In line $7$, KMeans method is utilized to make each row of $D$ into $Q$ clusters. Next, in line $8$, each $G_{f} \in \mathcal{G}$ will be allocated to the cluster to which the $f^{th}$ row of $D$ is assigned. Finally, in the line $9$, based on each $G_{f}$ of each cluster, we can obtain corresponding participating clients per cluster.
		
		\subsubsection{\textbf{Cluster-based Model Aggregation (CMA) in Each Round}}
		\label{Subsection_3_Cluster_Model_Aggregation}
		After determining $Q$ clusters in each round, we consider executing model aggregation (MA) in each cluster to generate cluster-based global model (CGM), where \texttt{FedAvg} \cite{H_B_McMahan_Communication_Efficient_Learning_Deep_Networks} algorithm is employed in this work. Let $q$ denote an obtained cluster that has a total of $N_q$ participating clients.  Hence, the CGM $\omega_{q}$ for cluster $q$ at round $t$ can be given as follows \cite{Zou_EFCKD}:
		\begin{subequations}\label{Opt_12}
			\vspace{-0.1cm}
			\setlength{\abovedisplayskip}{3pt}
			\setlength{\belowdisplayskip}{3pt}
			\begin{align}
				&\label{Opt_12:const1}
				\omega_{q} = \sum_{s=1}^{N_q}\frac{|\mathcal{X}_s|}{|\mathcal{X}_q|}\omega_s^t, \tag {12}
			\end{align}	
		\end{subequations}
		where $|\mathcal{X}_q| = \sum_{s=1}^{N_q}|\mathcal{X}_s|$ denotes the total number of images owned by cluster $q$. The generated CGM of a cluster will be sent to the ordinary nodes in that cluster (recognized by their IDs) to initialize local models to be trained in the next round.
		
		\subsubsection{\textbf{Model Aggregation in the Final Round for Future Application}}
		\par
		If only consider CMA, when new clients join, it is necessary to determine the proper cluster-based global model for those clients. If the number of clusters is large, selecting an appropriate global model will become cumbersome. Thus, we additionally consider model aggregation across all participating clients in the final round. The generated global model can be directly used by each newcomer without the necessity of global model selection process. As total number of participating clients in each round is denoted as $f$ before, the global model $\omega_g$ can be represented as below:   	
		\begin{subequations}\label{Opt_13}
			\vspace{-0.1cm}
			\setlength{\abovedisplayskip}{3pt}
			\setlength{\belowdisplayskip}{3pt}
			\begin{align}
				&\label{Opt_13:const1}
				\omega_g = \sum_{s=1}^{f}\frac{|\mathcal{X}_s|}{|\mathcal{X}|}\omega_s^t, \tag {13}
			\end{align}	
		\end{subequations}	
		where $|\mathcal{X}| = \sum_{s=1}^{f}|\mathcal{X}_s|$ indicates the number of images data of all participating clients.
		
		\begin{figure}[t!]
			\vspace{-0.25cm}
			\begin{algorithm}[H]	
				\renewcommand{\algorithmicrequire}{\textbf{Input:}}
				\renewcommand{\algorithmicensure}{\textbf{Output:}}
				\caption{Proposed OSC-FSKD Approach Towards Non-IID Satellite Imagery in Each Orbit}
				\label{alg_2}
					\begin{algorithmic}[1]
						\REQUIRE $\mathcal{X}_s = \{x_s^1, x_s^2, ..., x_s^M\}$, cluster number: $Q$
						\ENSURE Global Model $\omega_g$ \\
						\hspace{-0.55cm} \textbf{Process on the sink node:}
						\STATE Initialize the global model $\omega_g$	
						\STATE Distribute $\omega_g$ to each client for local model initialization  				
						\FOR{each round $t = 1, 2, ..., T$}
						\FOR{each participating ordinary node (i.e., $\beta_s=1$)}
						\IF {$t == 1$} 	    
						\STATE $\omega_s^t, \Delta\omega_s^t \leftarrow \textrm{ClientUpdate}(t=1)$
						\ELSE 
						\STATE $\omega_s^t, \Delta\omega_s^t \leftarrow \textrm{ClientUpdate}(t)$
						\ENDIF
						\ENDFOR
						\STATE Obtain $\mathcal{G}$ via \eqref{Opt_10}
						\STATE Get $Q$ clusters and determine clients of each cluster via \textbf{Algorithm \ref{alg_1}}
						\STATE Get each cluster-based global model $\omega_q$ via \eqref{Opt_12} 
						\STATE Submit each $\omega_q$ to each cluster $q$ to initialize the local models belong to each cluster $q$, respectively
						\IF {$t == T$} 
						\STATE Obtain global model $\omega_g$ via \eqref{Opt_13}
						\ENDIF
						\ENDFOR
						\STATE \textbf{return} $\omega_g$
						\hfill \break
						\\
						\hspace{-0.63cm} \textbf{ClientUpdate}(t):
						\FOR{each epoch}
						\FOR{batch $(x_s, y_s) \in \mathcal{D}_s$}
						\STATE Compute KD loss: $\mathcal{L}_s^\textrm{KD}$, if $t \neq 1$
						\STATE Compute CE loss: $\mathcal{L}_s^\textrm{CE}$
						\STATE Compute local loss via \eqref{Opt_11}: $\mathcal{L}_s^\textrm{loss}$
						\STATE $\omega_s \leftarrow \omega_s - \lambda\bigtriangledown\mathcal{L}_s^\textrm{loss}$ \cite{H_B_McMahan_Communication_Efficient_Learning_Deep_Networks}
						\ENDFOR
						\ENDFOR	
						\STATE Get the updated local model at round $t$: $\omega_s^t$
						\STATE Calculate the model update by calculating the difference between the local model before and after update : $\Delta\omega_s^t$
						\STATE \textbf{return} $\omega_s^t$, $\Delta\omega_s^t$
					\end{algorithmic}  
				\end{algorithm} 
				\vspace{-1cm}
			\end{figure}
			\subsubsection{\textbf{Algorithm Summary}}
			In this part, the proposed OSC-FSKD approach is summarized in Algorithm \ref{alg_2}. In Algorithm \ref{alg_2}, lines $1-19$ are regarding the process on the sink node, while lines $20-30$ are the process for each client. Specifically, we initialize the global model $\omega_g$ in line $1$ which will be used to initialize all the clients in line $2$. Then in each round (line $3$) for each participating ordinary node (line $4$), if it is in the first round (line $5$), the local model will be updated in line $6$ without considering SKD. Thus, the KD loss will not be computed to construct the local loss. Otherwise, the local model will be updated in line $8$, where SKD is adopted. Afterwards, $\mathcal{G}$ will be computed in line $11$, which will be used by the Algorithm $1$ to decide the clients for $Q$ clusters (line $12$). In line $13$, we calculate the cluster-based global model $\omega_q$ (CGM) for each cluster $q$ by \eqref{Opt_12}. Particularly, $\omega_q$ will be submitted to cluster $q$ to initialize the local models of cluster $q$ (line $14$). If current round $t$ is the final round $T$ (line $15$), then the global model for all participating clients will be generated via \eqref{Opt_13} in line $16$. From lines $20-30$, local model update will be performed. In particular, for each batch $(x_s, y_s) \in \mathcal{D}_s$ (line $21$) in each epoch (line $20$), the KD loss will be computed from the second round (line $22$) as SKD is considered to be performed from the second round, and then in line $23$, we compute the CE loss. Here, $\mathcal{D}_s = \{(x_s^1, y_s^1), (x_s^2, y_s^2), ..., (x_s^M, y_s^M)\}$ is the dataset owned by each ordinary node $s$ in orbit $k$ including image data and corresponding labels. Then we calculate the local loss in line $24$. In line $25$, for each batch in each epoch, the local model will be updated through $\omega_s \leftarrow \omega_s - \lambda\bigtriangledown\mathcal{L}_s^\textrm{loss}$ \cite{H_B_McMahan_Communication_Efficient_Learning_Deep_Networks}, where $\lambda$ is the learning rate. In line $28$, we get the updated local model at round $t$. The local model update $\Delta\omega_s^t$ will be obtained in line $29$. In line $30$, $\Delta\omega_s^t$ and $\omega_s^t$ will be returned to the sink node for cluster determination and model aggregation. 
			
			\section{Evaluation Result and Analysis}
			\label{sec_4_Evaluation_Result_Analysis}
			\subsection{Experimental Settings}
			\par
			As the proposed OSC-FSKD approach is considered to be applied in each orbit in this work, we consider evaluating the proposed OSC-FSKD in one orbit for simplicity. Specifically, we consider the number of ordinary nodes as $10$, $20$ and $30$, and one sink node for model aggregation in an orbit. As a case study, $3$ clusters are considered for simplicity. In addition, we consider $3$ satellite datasets: EuroSAT \cite{P_Helber_EuroSAT}, SAT4 \cite{S_Basu_DeepSat} and SAT6 \cite{S_Basu_DeepSat}. Specifically, EuroSAT is created for land use and land cover classification, which has $27,000$ labeled images belong to $10$ classes \cite{P_Helber_EuroSAT}. We split the whole EuroSAT dataset into two parts: $70\%$ data for training and the remaining $30\%$ for test. As per \cite{S_Basu_DeepSat}, SAT4 is a dataset that has $400,000$ and $100,000$ image patches for training and test, respectively. Those patches belong to $4$ classes: $\{$\texttt{barren land, grassland, trees, none}$\}$. Here, \texttt{none} means other land cover class. For simplicity, we randomly choose $10,000$ image patches from training part for training and $5,000$ image patches from test part for test in this work. SAT6 is a dataset includes $324,000$ images for training and $81,000$ for test, which covers $6$ classes: $\{$\texttt{barren land, grassland, trees, buildings, roads, water bodies}$\}$. For convenience, we randomly select $3,000$ images from training part and $1,000$ images from test part for evaluation. It is worth noting that we resize each image to $224 \times 224$ pixels. To prepare the non-IID data for each ordinary node, \textit{Dirichlet Distribution} with parameter $Dir = \{0.6, 0.9, 5\}$ is adopted. 
			
			\par
			To construct each local client, we employ two network backbones: CNN and Swin Transformer \cite{Z_Liu_Swin_Transformer}. Particularly, regarding CNN, we consider two convolutional layers as representation layers and one fully connected (FC) layer as a classifier. Towards EuroSAT dataset, the input channel for two convolutional layers is $\{3, 64\}$, while the output channel is $\{64, 64\}$, respectively. In terms of the FC layer, the input is $64*224*224$, while the output is the number of classes (i.e. $10$). For SAT4 and SAT6, since each image contains $4$ bands (i.e., R, G, B, and NIR) \cite{N_Yang_DropBand_CNN_VHR}, the input of the first convolutional layer is $4$, and other setting is the same as EuroSAT. For Swin Transformer, Swin-T \cite{Z_Liu_Swin_Transformer} is adopted. Parameters used by both CNN and Swin-T are listed as below: we set the learning rate as $0.001$, batch size as $32$, local epoch as $10$, temperature $\tau$ as $0.1$ \cite{T_Liang_Compressing_Multiobject_Tracking_Model} and the hyperparametrer $\varphi$ as $0.9$. For the weight parameters in the beginning, random initialization \cite{B_Salehi_FLASH_FL_Automated_Selection} is adopted. Besides, we consider a total of $40$ rounds to train the EuroSAT dataset, while $30$ rounds to train both the SAT4 and SAT6 datasets. Apart from this, SGD is adopted as the optimizer.
			\par
			For comparison, the following baselines are adopted. 1) \texttt{FedProx} \cite{FedProx}: a generalization of FedAvg, where the proximal term is added to the local subproblem. 2) \texttt{FedAU} \cite{FedAU}: a method aims to improve FedAvg by considering to weight the client updates adaptively via online optimal weights estimation. 3) \texttt{FedALA} \cite{FedALA}: a FL method with an adaptive local aggregation module. 4) \texttt{FedSwin}: a method that we combine FL with Swin-T \cite{Z_Liu_Swin_Transformer}. 5) \texttt{pFedSD} \cite{pFedSD}: a SKD-based personalized FL approach. Particularly, for \texttt{FedProx}, \texttt{FedAU} and \texttt{FedALA}, we adopt Swin-T to build the local client. For \texttt{pFedSD}, same as the proposed method, we consider adopting CNN and Swin-T to construct each local client, respectively. To distinguish them, we use \texttt{pFedSD} to denote the one with Swin-T, and use \texttt{pFedSD$^*$} to represent the one with CNN.
			\par
			To evaluate the final obtained global model for achieving \eqref{Opt_6}, except from calculating the \textit{observation accuracy (OA)} of the participating ordinary nodes in a single orbit, other two metrics are also employed: 1) \textit{average accuracy (AA)}, and 2) \textit{Macro F1-Score} \cite{Weinfeld_Document_Classification_Integrating}. Specifically, to obtain AA, we first calculate the accuracy for each client, then take the average. For Macro F1-Score, we first compute the F1-Score per class, then get the average value with equal weight per class \cite{Weinfeld_Document_Classification_Integrating}. 
			
			\vspace{-0.1cm}
			\subsection{Numerical Analysis}	

			\par
			In Table \ref{sec_4_tab1}, we show the macro F1-Score and observation accuracy (OA) achieved by the proposed OSC-FSKD approach for SAT4 and EuroSAT datasets with $Dir=0.9$, where CNN and Swin-T are respectively used as the backbone network for each local client. Through this table, it can be observed that the proposed method with Swin-T as backbone can achieve higher Macro F1-Score and OA for both SAT4 and EuroSAT datasets. Thus, it can be inferred that using Swin-T as backbone is better than the considered CNN in this work. For simplicity, next, we will utilize Swin-T as backbone for evaluation.
			\begin{table}[t!]
				\vspace{-0.05cm}
				\caption{Macro F1-Score and OA (i.e., \eqref{Opt_6}) achieved by the proposed OSC-FSKD with two backbones for local clients.}
				\vspace{-0.2cm}
				\renewcommand\arraystretch{1}
				\begin{center}
					\begin{tabular}{|c|m{1.6cm}||m{2cm}|m{1.6cm}|}
						\hline
						\hfil \textbf{Dataset} & \hfil \textbf{Backbone} & \hfil \textbf{Macro F1-Score
						} & \hfil \textbf{OA} \\ \hline
						\hfil \multirowcell{2}{SAT4}& \hfil CNN & \hfil $0.9263$ & \hfil $0.9304$  \\ \cline{2-4}
						& \hfil Swin-T & \hfil $0.9619$ & \hfil $0.9650$ \\ 	
						\hline \hline	
						\multirowcell{2}{EuroSAT}& \hfil CNN & \hfil $0.5833$ & \hfil $0.5964$  \\ \cline{2-4}	
						& \hfil Swin-T & \hfil $0.8942$ & \hfil $0.8983$ \\ \hline							
					\end{tabular}
					\label{sec_4_tab1}
				\end{center}
				\vspace{-0.7cm}
			\end{table}
			
			\begin{figure} [t!]
				\centering
				\vspace{-0.1cm}
				\includegraphics[scale = .4]{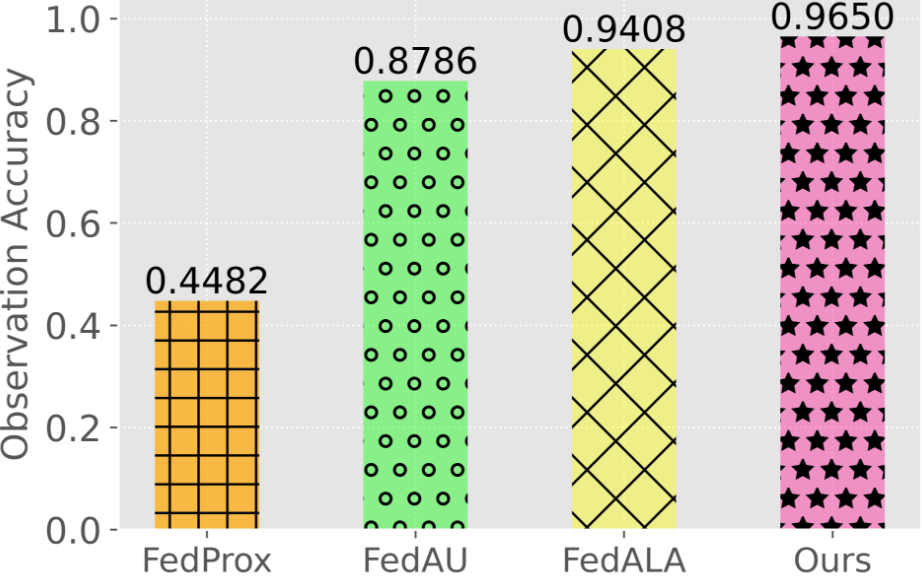}
				\vspace{-0.1cm}
				\caption{Accuracy comparison among FedProx, FedAU, FedALA and Ours (Swin-T) using SAT4 dataset (Test).}
				\label{fig_accuracy_FedProx_FedAU_FedALA_Ours} 
				\vspace{-0.3cm}
			\end{figure}
			\par
			In Fig. \ref{fig_accuracy_FedProx_FedAU_FedALA_Ours}, we compare the proposed OSC-FSKD approach with FedProx, FedAU and FedALA approach in terms of observation accuracy (OA) by using SAT4 dataset ($Dir=0.9$). It can be seen that the OA achieved by the proposed method is $2.15\times$, $1.10\times$ and $1.03\times$ greater than FedProx, FedAU and FedALA, respectively. The reason is: among those methods, only the proposed method is designed based on the clustered federated learning, which is possible to solve the dilemma of performance degradation caused by learning a shared global model when the data is heterogeneous \cite{C_Li_FL_Soft_CLustering}. Accordingly, the proposed method outperforms the other three baselines.

			
			\begin{figure}[t!]
				\centering
				\begin{subfigure}[b]{0.24\textwidth} 
					\includegraphics[width=1 \linewidth]{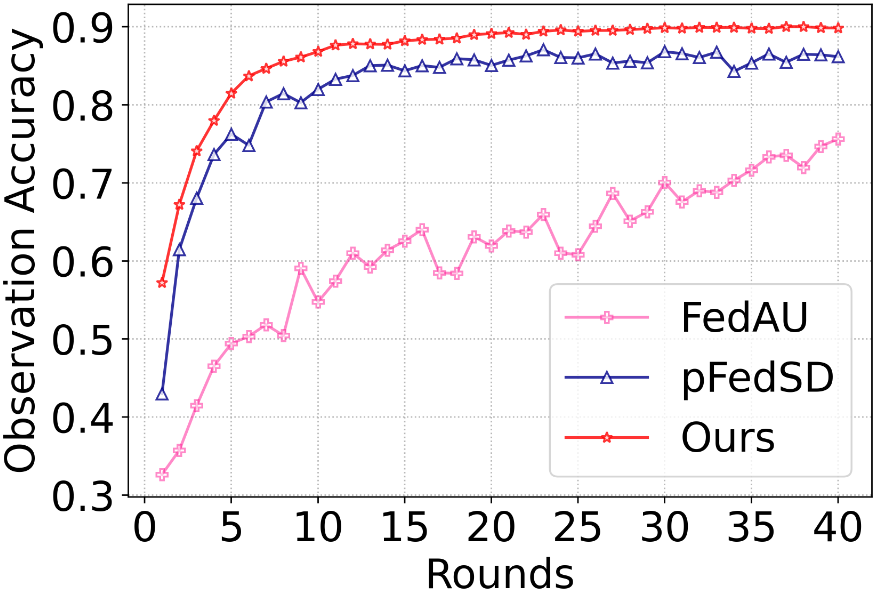}
					\captionsetup{font=small}
					\caption{EuroSAT}
					\label{Comparison_Among_FedAU_pFedSD_Ours_EuroSAT}
				\end{subfigure} 
				\begin{subfigure}[b]{0.24\textwidth} 
					\includegraphics[width=1 \linewidth]{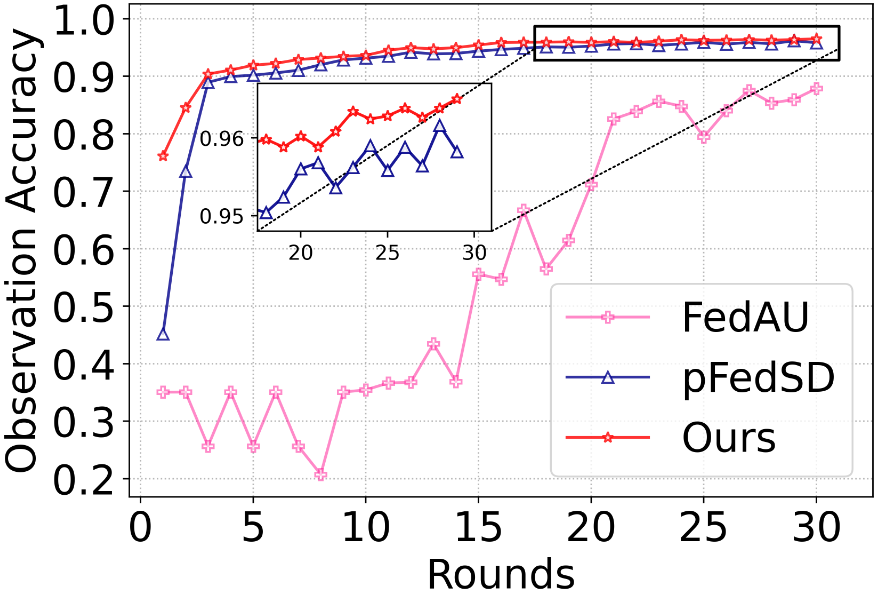}
					\captionsetup{font=small}
					\caption{SAT4}
					\label{Comparison_Among_FedAU_pFedSD_Ours_SAT4}
				\end{subfigure}
				\vspace{-0.4cm}
				\captionsetup{font=small}
				\caption {Comparison among FedAU, pFedSD and the proposed method (Swin-T) using EuroSAT and SAT4 datasets (Test).}
				\label{sec_4_Comparison_Among_CNN_KD_Ours}
				\vspace{-0.55cm}
			\end{figure}
			
			\begin{table*}[t!]
				\caption{Comparison among FedSwin, pFedSD${^*}$, pFedSD, and the proposed method in terms of average accuracy (AA), observation accuracy (OA, \eqref{Opt_6}) and Macro F1-Score. The best results are shown in bold.}
				\vspace{-0.1cm}
				\makebox[\textwidth][c]{
					\begin{tabular}{m{1.6cm}|m{1.7cm}|m{1.5cm}m{1.5cm}m{1.5cm}|m{1.7cm} m{1.7cm} m{2cm}}
						\toprule
						\hfil  Dataset	& \hfil Method  & \hfil SKD & \hfil Swin-T & \hfil Clustering & \hfil AA
						& \hfil OA & \hfil Macro F1-Score
						\\
						\hline
						\midrule
						\hfil	\multirow{4}{*}{SAT4} & \hfil FedSwin & \hfil \faTimes  & \hfil \faCheck & \hfil \faTimes  & \hfil $0.9502$ & \hfil $0.9534$ & \hfil $0.9480$
						\\ \cline{2-8} 
						& \hfil pFedSD${^*}$ & \hfil \faCheck & \hfil \faTimes & \hfil \faTimes  & \hfil $0.7366$ & \hfil $0.7264$ & \hfil $0.7219$ \\ \cline{2-8}
						& \hfil pFedSD & \hfil \faCheck & \hfil \faCheck &  \hfil \faTimes & \hfil $0.9556$ & \hfil $0.9582$ & \hfil $0.9553$ \\ \cline{2-8}
						&  Ours (Swin-T) & \hfil \faCheck & \hfil \faCheck & \hfil \faCheck & \hfil $\textbf{0.9617}$ & \hfil $\textbf{0.9650}$ & \hfil $\textbf{0.9619}$ \\ \hline
						\midrule
						\hfil \multirow{4}{*}{SAT6} & \hfil FedSwin & \hfil \faTimes & \hfil \faCheck & \hfil \faTimes & \hfil $0.8861$ & \hfil $0.9040$ & \hfil $0.7451$\\ \cline{2-8}
						& \hfil pFedSD${^*}$ & \hfil \faCheck & \hfil \faTimes & \hfil \faTimes  & \hfil $0.7793$ & \hfil $0.7860$ & \hfil $0.6299$ \\ \cline{2-8}
						& \hfil pFedSD & \hfil \faCheck & \hfil \faCheck & \hfil \faTimes & \hfil $0.9028$  & \hfil $0.9100$ & \hfil $0.8407$ \\ \cline{2-8}
						& \hfil Ours (Swin-T) & \hfil \faCheck & \hfil \faCheck & \hfil \faCheck & \hfil $\textbf{0.9436}$ & \hfil $\textbf{0.9480}$ & \hfil $\textbf{0.8774}$ \\ 
						\hline
						\midrule
						\hfil	\multirow{4}{*}{EuroSAT} & \hfil FedSwin & \hfil \faTimes & \hfil \faCheck & \hfil \faTimes & \hfil $0.8699$ & \hfil $0.8714$ & \hfil $0.8673$ \\ \cline{2-8}
						& \hfil pFedSD${^*}$ & \hfil \faCheck & \hfil \faTimes & \hfil \faTimes  & \hfil $0.4412$ & \hfil $0.4399$ & \hfil $0.4370$ \\ \cline{2-8}
						& \hfil pFedSD & \hfil \faCheck & \hfil \faCheck & \hfil \faTimes  & \hfil $0.8624$ & \hfil $0.8616$ & \hfil $0.8602$ \\ \cline{2-8}
						& \hfil Ours (Swin-T) & \hfil \faCheck & \hfil \faCheck & \hfil \faCheck & \hfil $\textbf{0.8981}$  & \hfil $\textbf{0.8942}$ & \hfil $\textbf{0.8983}$ \\ 
						\bottomrule
					\end{tabular}%
				}
				\label{sec_4_tab2}%
				\vspace{-0.38cm}
			\end{table*}%
			
			\par
			In Fig. \ref{sec_4_Comparison_Among_CNN_KD_Ours}, we compare the proposed method with FedAU and pFedSD regarding the observation accuracy (OA) across various rounds for EuroSAT and SAT4 datasets ($Dir=0.9$). It can be seen that the proposed method exhibits a more significant advantage in OA of those two datasets than FedAU. Compared with pFedSD, the effectiveness of the proposed method is more noticeable than pFedSD over the EuroSAT regarding OA in each round. For the SAT4 dataset, the OA achieved by the proposed method is slightly higher than that of pFedSD when gradually achieving convergence. This is because pFedSD and FedAU only learn a single shared global model. This will decrease the learning performance due to the existence of non-IID data as per \cite{C_Li_FL_Soft_CLustering}.
			

			\par
			In Table \ref{sec_4_tab2}, the comparison among FedSwin, pFedSD*, pFedSD, and the proposed method regarding average accuracy (AA), OA (i.e., \eqref{Opt_6}), and Macro F1-Score is conducted, where three datasets are adopted and $Dir=0.9$. As we can see from Table \ref{sec_4_tab2}, only the proposed method adopt SKD, Swin-T and clustering. For SAT4 dataset, the AA, OA, and Macro F1-Score realized by the proposed method separately are $0.9617$, $0.9650$, and $0.9619$, which are the highest values compared with other baselines. Interestingly, the same phenomenon can also be found for SAT6 and EuroSAT datasets. Thus, the proposed method is the best among those methods.
			\begin{table}[t!]
				\caption{OA (i.e., \eqref{Opt_6}) achieved by considering various heterogeneous levels for SAT4 dataset (Test).}
				\vspace{-0.3cm}
				\renewcommand\arraystretch{1}
				\begin{center}
					\begin{tabular}{p{0.11\textwidth}>{\centering}p{0.08\textwidth}>{\centering}p{0.08\textwidth}>{\centering\arraybackslash}p{0.08\textwidth}}
						\hline
						\hfil \multirow{2}{*}{\textbf{Method}}&\multicolumn{3}{c}{\textbf{Dirichlet Distribution}}\\\cline{2-4}
						& $\textbf{Dir(0.6)}$ & $\textbf{Dir(0.9)}$ & $\textbf{Dir(5)}$\\
						\hline
						\hfil FedAU & $0.6410$ & $0.8786$ & $0.9132$ \\
						\hfil FedALA & $0.9222$ & $0.9408$ & $0.9306$ \\
						\hfil Ours (Swin-T) & $\textbf{0.9628}$ & $\textbf{0.9650}$ & $\textbf{0.9652}$ \\
						\hline
					\end{tabular}
					\label{sec_4_tab3}
				\end{center}
				\vspace{-0.1cm}
			\end{table}
			\par 
			In Table \ref{sec_4_tab3}, we demonstrate the performance with different heterogeneity levels, where we adopt $Dir = \{0.6, 0.9, 5\}$ for Dirichlet Distribution. The lower value the higher the data heterogeneity level. It can be observed that although the OA (i.e., \eqref{Opt_6}) obtained by the proposed method decreases slightly when the heterogeneity increases, the proposed method still achieves the highest OA compared with FedAU and FedALA. Hence, we can infer that the proposed method outperforms the other two methods when dealing with heterogeneous data.
			
			\begin{table}[t!]
				\caption{OA (i.e., \eqref{Opt_6}) achieved by considering different number of clients for SAT4 dataset (Test).}
				\vspace{-0.25cm}
				\renewcommand\arraystretch{1}
				\begin{center}
					\begin{tabular}{p{0.11\textwidth}>{\centering}p{0.08\textwidth}>{\centering}p{0.08\textwidth}>{\centering\arraybackslash}p{0.08\textwidth}}
						\hline
						\hfil \multirow{2}{*}{\textbf{Method}}&\multicolumn{3}{c}{\#\textbf{Clients}} \\\cline{2-4}
						& =$10$ & =$20$ & =$30$ \\
						\hline
						\hfil FedAU & $0.8786$ & $0.7449$ & $0.4643$ \\
						\hfil FedALA & $0.9408$ & $0.9116$ & $0.9034$ \\
						\hfil Ours (Swin-T) & $\textbf{0.9650}$ & $\textbf{0.9355}$ & $\textbf{0.9298}$ \\
						\hline
					\end{tabular}
					\label{sec_4_tab4}
				\end{center}
				\vspace{-0.65cm}
			\end{table}
			\par
			In Table \ref{sec_4_tab4}, we evaluate the scalability of the proposed method when considering various number of clients. Compared with FedAU and FedALA, we can see that the proposed method can achieve the highest OA (i.e., \eqref{Opt_6}) regardless of the number of clients. Therefore, it can be conjectured that the proposed method is superior to the other two methods.


			\begin{figure}[h]
				\vspace{-0.15cm}
				\centering
				\begin{subfigure}{0.4\textwidth} 
					\includegraphics[width=1.09 \linewidth]{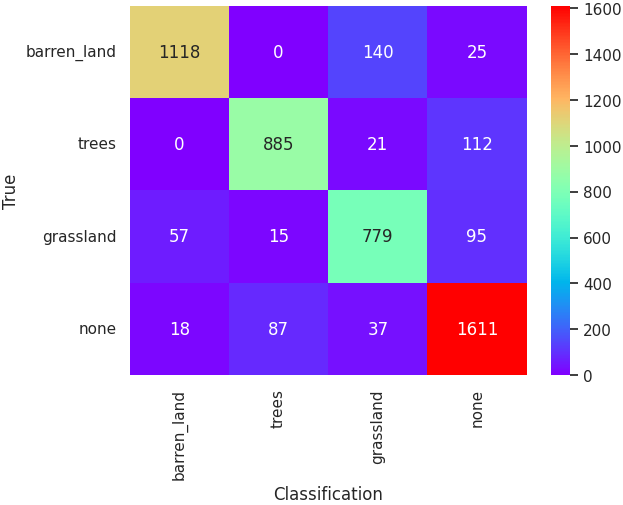}
					\captionsetup{font=small} 
					\caption{FedAU}
					\label{FedAU_CM_SAT4}
				\end{subfigure} \hfil \\
				\begin{subfigure}{0.43\textwidth} 
					\includegraphics[width=1.05 \linewidth]{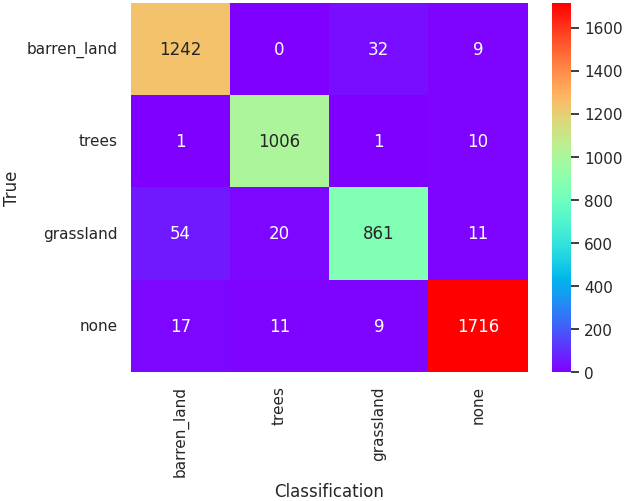}
					\captionsetup{font=small} 
					\caption{Ours (Swin-T)}
					\label{Ours_CM_SAT4}
				\end{subfigure} \hfil 
				
				\vspace{-0.1cm}
				\captionsetup{font=small} 
				\caption {Confusion matrix achieved by FedAU and the proposed method via using the test dataset of SAT4 dataset.}
				\label{Confusion_Matrix_Multi_Class_Classification}
				\vspace{-0.7cm}
			\end{figure}
			\par
			In Fig. \ref{Confusion_Matrix_Multi_Class_Classification}, we show the confusion matrix achieved by FedAU and the proposed method over the SAT4 dataset. Through this figure, we can know the total correctly detected number achieved by FedAU is $1, 118 + 885 + 779 + 1, 611 = 4, 393$, while that achieved by the proposed method is $4, 825$. Therefore, the effectiveness of the proposed method in achieving correct observation is better than that of FedAU.
			
			
			\vspace{-0.1cm}
			\section{Conclusion}
			\label{sec_5_conclusion}
			In this paper, an orbit-based spectral clustering-assisted clustered federated self-knowledge distillation approach has been proposed for Earth observation in each orbit, aiming to maximize the observation accuracy for the ordinary nodes in a single orbit. Specifically, normalized Laplacian-based spectral clustering is employed in each round to group clients based on the cosine similarity computed by mode updates. Then the model aggregation for each cluster will be executed in each round, and model aggregation will be performed across all participating clients in the final round for the ease of future application. In addition, self-knowledge distillation is executed from the second round, in which the knowledge from the local model obtained in the previous round will be distilled to guide the current local model training. For evaluation, EuroSAT, SAT4, and SAT6 datasets are used, meanwhile the Dirichlet Distribution is introduced to prepare the non-IID data for all the ordinary nodes. The evaluation demonstrates that the proposed method can obtain remarkable results in terms of observation accuracy, average accuracy, and Macro F1-Score. For instance, consider Dirichlet Distribution parameter is $0.9$, the observation accuracy achieved by the proposed method outperforms FedSwin, pFedSD, FedAU, and FedALA by $1.16\%$, $0.68\%$, $8.64\%$, and $2.42\%$, respectively.
			

		\end{document}